\documentclass[
    reprint,
    superscriptaddress,
    amsmath,amssymb,
    aps,nofootinbib
]{revtex4-2}

\usepackage{graphicx}
\usepackage{amsmath}
\usepackage{amssymb}
\usepackage{float}
\usepackage[dvipsnames]{xcolor}
\usepackage[colorlinks=true,urlcolor=blue,linkcolor=blue,citecolor=blue, hypertexnames=false]{hyperref}
\usepackage{cleveref}
\usepackage{physics}
\usepackage[detect-weight=true,separate-uncertainty=true]{siunitx}

\crefname{equation}{Eq.}{Eqs.}
\Crefname{equation}{Equation}{Equations}
\crefname{figure}{Fig.}{Figs.}
\Crefname{figure}{Figure}{Figures}
\crefname{section}{Sec.}{Sects.}
\Crefname{section}{Section}{Sections}
\crefname{table}{Table}{Tables}
\crefname{appendix}{Appendix}{Apps.}
\Crefname{appendix}{Appendix}{Apps.}

\newcommand{\aop}{\hat a}

\def\mean#1{\langle #1 \rangle}

\newcommand{\captiontitle}[1]{#1}

\begin{document}

\author{Crist\'obal Lled\'o$^\dag$}
\email{cristobal.lledo.veloso@usherbrooke.ca}
\affiliation{Institut Quantique and D\'epartement de Physique, Universit\'e de Sherbrooke, Sherbrooke J1K 2R1 QC, Canada}
\author{R\'emy Dassonneville$^\dag$}
\affiliation{Ecole Normale Sup\'erieure de Lyon, CNRS, Laboratoire de Physique, F-69342 Lyon, France}
\author{Adrien Moulinas}
\affiliation{Institut Quantique and D\'epartement de Physique, Universit\'e de Sherbrooke, Sherbrooke J1K 2R1 QC, Canada}
\author{Joachim Cohen}
\affiliation{Institut Quantique and D\'epartement de Physique, Universit\'e de Sherbrooke, Sherbrooke J1K 2R1 QC, Canada}
\author{Ross Shillito}
\affiliation{Institut Quantique and D\'epartement de Physique, Universit\'e de Sherbrooke, Sherbrooke J1K 2R1 QC, Canada}
\author{Audrey Bienfait}
\affiliation{Ecole Normale Sup\'erieure de Lyon,  CNRS, Laboratoire de Physique, F-69342 Lyon, France}
\author{Benjamin Huard}
\affiliation{Ecole Normale Sup\'erieure de Lyon,  CNRS, Laboratoire de Physique, F-69342 Lyon, France}
\author{Alexandre Blais}
\affiliation{Institut Quantique and D\'epartement de Physique, Universit\'e de Sherbrooke, Sherbrooke J1K 2R1 QC, Canada}
\affiliation{Canadian Institute for Advanced Research, Toronto, M5G1M1 Ontario, Canada}

{\let\thefootnote\relax\footnote{{${}^\dag$ These authors contributed equally to this work.}}}

\title{Cloaking a qubit in a cavity}

\date{\today}

\begin{abstract}
Cavity quantum electrodynamics (QED) uses a cavity to engineer the mode structure of the vacuum electromagnetic field such as to enhance the interaction between light and matter. Exploiting these ideas in solid-state systems has lead to circuit QED which has emerged as a valuable tool to explore the rich physics of quantum optics and as a platform for quantum computation. 
Here we introduce a simple approach to further engineer the light-matter interaction in a driven cavity by controllably decoupling a qubit from the cavity's photon population, effectively cloaking the qubit from the cavity. This is realized by driving the qubit with an external tone tailored to destructively interfere with the cavity field, leaving the qubit to interact with a cavity which appears to be in the vacuum state. Our experiment demonstrates how qubit cloaking can be exploited to cancel the ac-Stark shift and measurement-induced dephasing, and to accelerate qubit readout. In addition to qubit readout, applications of this method include qubit logical operations and the preparation of non-classical cavity states in circuit QED and other cavity-based setups. 
\end{abstract}

\maketitle

\section*{Introduction}

Cavity and circuit QED explore light-matter interaction at its most fundamental level, providing the tools to control the dynamical evolution of single atoms and photons in a deterministic fashion~\cite{Haroche2006}. This has allowed circuit QED to emerge as a platform to explore the rich physics of quantum optics in novel parameter regimes and to become a leading architecture for quantum computing~\cite{Blais2021}. In this system, strong drives on the cavity are used to realize multi-qubit gates~\cite{Paik2016,Lu2022}, to stabilize quantum states of the cavity~\cite{Leghtas2015,Gao2018,Grimm2020,Berdou2023}, and for qubit readout~\cite{Blais2004, Blais2021}. However, even under moderate cavity photon populations, cavity drives often lead to undesired effects such as qubit transitions~\cite{Minev2019} resulting in reduced readout fidelity~\cite{Walter2017}, increased dephasing~\cite{Lu2022,Grimm2020}, and imperfect quantum state stabilization~\cite{Leghtas2015,Gao2018,Berdou2023}. Other consequences of cavity drives in the dispersive qubit-cavity regime are the qubit ac-Stark shift, which can result in unwanted phase accumulations~\cite{Eickbusch2022}, measurement-induced dephasing~\cite{Schuster2005,Gambetta2006}, and Kerr nonlinearity~\cite{Kirchmair2013}.

\begin{figure}[t!]
\includegraphics[width=0.8\columnwidth]{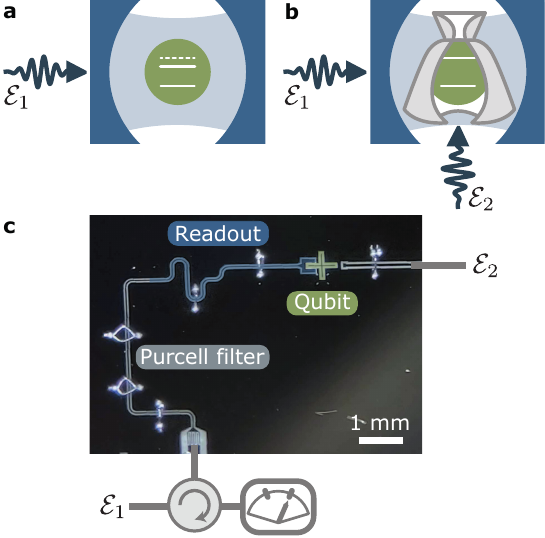}
\caption{\captiontitle{Concept and device.} \textbf{a}, Schematic illustration of a qubit (green) coupled to a driven cavity represented by two mirrors (blue). A drive $\mathcal{E}_1$ on the cavity displaces the cavity field, which effectively acts as a classical drive on the qubit. This results in qubit ac-Stark shift and measurement-induced dephasing (i.e.~level broadening), see the full white lines. \textbf{b}, A second drive $\mathcal{E}_2$ of appropriate time-dependent amplitude, frequency and phase cloaks the qubit from the effective classical field resulting from $\mathcal{E}_1$ interferometrically cancelling the ac-Stark shift and broadening. \textbf{c}, Optical image of the device which includes a transmon qubit (green), a readout cavity (dark blue) and a Purcell filter (gray).}
\label{fig: cartoon}
\end{figure}

In this work, we introduce a simple approach to engineer light-matter interaction in a driven cavity and prevent some of these unwanted effects. We show how an appropriately tailored drive on the qubit can decouple the qubit state and the cavity's photon population from one another, resulting in both systems interacting only through vacuum fluctuations of the cavity field. This qubit cloaking mechanism can be exploited to prepare non-classical states of the cavity field. Moreover, in the dispersive qubit-cavity regime, it results in the absence of ac-Stark shift and measurement-induced dephasing. This observation can be used to apply logical operations on the qubit in the presence of a cavity photon population, something which we exploit to accelerate qubit readout. Here, we experimentally demonstrate qubit cloaking using a transmon qubit~\cite{Koch2007} coupled to a coplanar waveguide resonator (\cref{fig: cartoon}.c).

\section*{Results}

When loaded with a coherent state, the cavity field acts as an effective classical drive on the qubit. A simple intuition behind qubit cloaking is that an additional drive on the qubit can be designed to interfere destructively with this effective drive, resulting in an empty cavity from the perspective of the qubit. As an illustration of this concept, consider a transmon qubit coupled to a microwave cavity, see \cref{fig: cartoon}a,b for a schematic representation. Without driving, the system Hamiltonian reads~\cite{Blais2021}
\begin{equation}\label{eq: H0}
        \hat H_\mathrm{0} =  
        4 E_C \hat n_\mathrm{tr}^2 -E_J \cos\hat \varphi_\mathrm{tr} + \hbar\omega_r \hat a^\dag \hat a
        + i\hbar g\hat n_\mathrm{tr} (\hat a^\dag - \hat a),
\end{equation}
with $\hat n_\mathrm{tr}$ and $\hat \varphi_\mathrm{tr}$ the transmon charge and phase operators, $E_C$ the charging energy, $E_J$ the Josephson energy, $\hat a^{(\dag)}$ the annihilation (creation) operator of the cavity mode of frequency $\omega_r$, and $g$ the transverse coupling rate. The cavity drive is $\hat H_1 = i \hbar \mathcal{E}_1(t) (\hat a^\dag - \hat a)$, where $\mathcal{E}_1(t)= \varepsilon_1(t) \sin(\omega_1 t + \phi_1)$ is the amplitude of the cavity drive at frequency $\omega_1$ with envelope $\varepsilon_1(t)$. The cloaking is triggered by a cancelling drive $\mathcal{E}_2(t)$ on the transmon qubit, which leads to an additional term $\hat H_\mathrm{2} = \hbar\mathcal{E}_2(t) \hat n_\mathrm{tr}$, such that the total system Hamiltonian is $\hat H = \hat H_0 + \hat H_\mathrm{1} + \hat H_\mathrm{2}$. 

The appropriate choice of $\mathcal{E}_2(t)$ that cloaks the qubit from the cavity field is revealed by moving to a displaced frame using the transformation $\hat D(\alpha_t) = \exp(\alpha_t\hat a^\dag - \alpha_t^* \hat a)$ under which $\hat D^\dag(\alpha_t) \aop \hat D(\alpha_t) = \aop + \alpha_t$. Choosing $\alpha_t = \int_0^t d\tau \mathcal E_1(\tau) e^{-i\omega_r(t-\tau)}$ has the effect of cancelling the cavity drive $\hat H_1$ in the displaced Hamiltonian, resulting in an effectice drive on the qubit of the form $ig(\alpha_t^* - \alpha_t) \hat n_\mathrm{tr}$ owing to the qubit-cavity coupling. Therefore, taking the amplitude of the cancelling tone to be the opposite,
\begin{equation}\label{eq: E2}
    \mathcal{E}_2(t) = -ig[\alpha_t^*-\alpha_t],
\end{equation}
disables any drive term in the displaced frame, where the total Hamiltonian comes down to (see Supplementary Note 1 for more details~\cite{sm})
\begin{equation} \label{eq: Displaced Hamiltonian}
        \hat H' 
        = \hat D^\dag(\alpha_t) \hat H \hat D(\alpha_t) - i \hat D^\dag(\alpha_t) \dot{\hat D}(\alpha_t)
        =\hat H_0.
\end{equation}
In short, despite the presence of the drive $\mathcal{E}_1$ populating the cavity with an average photon number $|\alpha_t|^2$, the qubit experiences the cavity as if it was in the vacuum state, only coupling to vacuum fluctuations of the cavity field.  Note that no approximations have been made to arrive at this result which is valid irrespective of the qubit-cavity detuning and for arbitrary time-dependent drive amplitude $\varepsilon_1(t)$ and frequency $\omega_{1}$. Beyond the transmon qubit, this result is valid for two-level systems and any nonlinear system linearly coupled to a cavity. The results are unchanged in the presence of qubit decay or dephasing, and the above derivation can exactly account for the finite decay rate $\kappa$ of the cavity by making the change $\omega_r \rightarrow \omega_r - i\kappa/2$ in the expression for $\alpha_t$. In the rotating-wave approximation and for a constant $\varepsilon_1$, the amplitude $\alpha_t$ entering the expression for $\mathcal{E}_2$ in \cref{eq: E2} is $(\varepsilon_1e^{-i(\omega_1 t + \phi_1)}/2)/(\omega_r-\omega_1-i\kappa/2)$ in steady state. See Supplementary Note 1~\cite{sm} for details.

This approach relies on two distinct driving ports---one port dedicated to exciting the qubit or nonlinear mode and a second port for driving the cavity mode---and is directly applicable in several cavity-based platforms including semiconducting quantum dots coupled to microwave resonators~\cite{Mi2017, Landig2018, Stockklauser2017, Samkharadze2018}, electrons on solid neon~\cite{Zhou2022}, circuit quantum acoustodynamics~\cite{Manenti2017}, and trapped atoms in cavity QED~\cite{Ye1999}. While it is an exact and robust result, potential limitations of qubit cloaking are discussed in the Supplemental Note 3~\cite{sm}.

\begin{figure}[t!]
\includegraphics{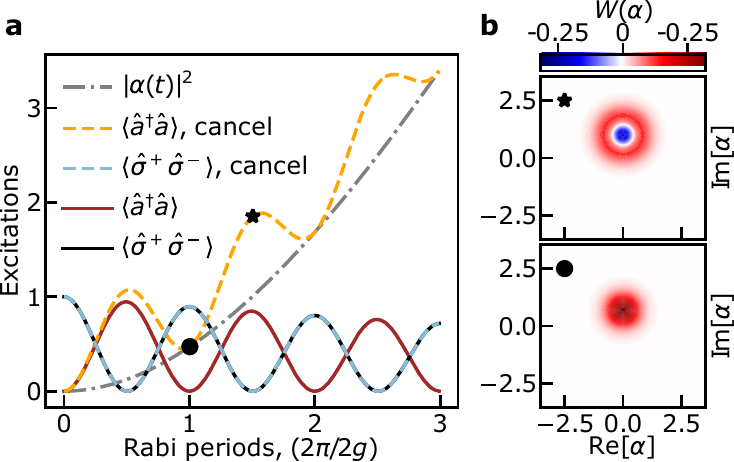}
\caption{\captiontitle{Vacuum Rabi oscillations in a filled cavity}.
\textbf{a}, Red and black solid lines correspond to damped vacuum Rabi oscillations of the resonant cavity and qubit, respectively, in the absence of any drive. The qubit-cavity Jaynes-Cummings coupling rate $g'$ is set to 100/7 of the cavity decay rate $\kappa$. With a cavity drive $\epsilon_1=46\kappa/7$ and the cancellation turned on, the cavity field (dashed orange) oscillates on top of $|\alpha_t|^2$ (dashed-dotted gray), swapping a single quantum of excitation back and forth with the qubit (dashed light blue). \textbf{b},~Computed Wigner distributions of the cavity field at 3/2 and 1 Rabi periods with the drives on, as indicated by the symbols in \textbf{a}. At 1 Rabi period, the cavity is in a coherent state, while it is in a displaced Fock state at 3/2 Rabi periods. 
}
\label{fig: vacuum Rabi}
\end{figure}

As a first example, we consider a situation where the qubit is resonant with the cavity, and model the transmon as a two-level system. In the absence of any drives, initializing the qubit in its first excited state and the cavity in the vacuum state leads to vacuum Rabi oscillations at the frequency $2g'$~\cite{Haroche2006}, where $g'=(g/2)(E_J/2E_C)^{1/4}$ is the Jaynes-Cummings coupling~\cite{Blais2021}. This is illustrated in \cref{fig: vacuum Rabi}a which shows the results (full black and red lines) of integration of the system's master equation including cavity decay, see Supplementary Note 1 for details~\cite{sm}. With the cavity drive and the cancelling tone present, the cavity population (orange dashed lines) increases following $|\alpha_t|^2$ (dashed-dotted grey line) on top of which oscillations are observed. On the other hand, the qubit population (light blue dashed line) is identical to that observed in the absence of the drives. In other words, instead of collapse and revival which are expected in the presence of a coherent state in the cavity~\cite{Brune1996,Eberly1980}, here the qubit undergoes oscillations at the vacuum Rabi frequency. The same conclusion holds when initializing the cavity mode in the Fock state $\ket1$ and the qubit in the ground state (not shown). This is a clear illustration that under cloaking the qubit only couples to vacuum fluctuations of the cavity. In related work, \textcite{Alsing1992} have shown how a drive on the atom can suppress atomic fluorescence in the steady-state of a lossless cavity. In contrast, cloaking is an exact result which is valid at all times and in the presence of loss. Moreover, this approach can be used to prepare non-classical states of the cavity field. For example, displaced Fock states $\hat D(\alpha) \ket{n=1}$ with arbitrary $\alpha$ can be prepared by starting with the same initial state as in \cref{fig: vacuum Rabi}a, and waiting for half-integer Rabi periods under appropriate cavity drive amplitude and phase. The Wigner function of the displaced Fock state is plotted at 3/2 Rabi periods in \cref{fig: vacuum Rabi}b, and is compared to that of the coherent state at 1 Rabi period. 

\begin{figure}[t!]
\includegraphics[scale=1]{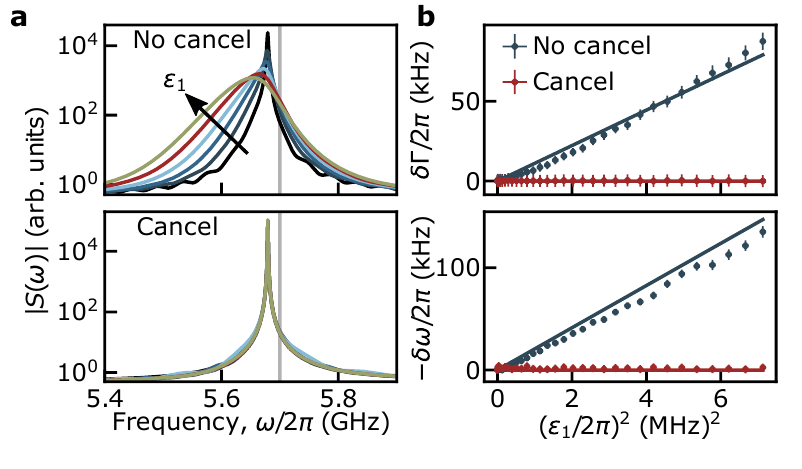}
\caption{\captiontitle{Cancellation of the ac-Stark shift and measurement-induced dephasing.} \textbf{a}, Numerical simulation of a two-level qubit spectral density without cancellation (top) and with cancellation (bottom) for various drive amplitudes $\varepsilon_1/2\pi$ from 10 to 60~MHz. The simulation parameters are $\omega_q/2\pi = 5.7$ GHz, $\omega_{r,1}/2\pi =7.6$ GHz, $g'/2\pi = 200$ MHz, and $\kappa/2\pi = 50$ MHz. \textbf{b}, Dots: Experimentally measured increased dephasing rate $\delta \Gamma$ (top) and qubit frequency shift $\delta\omega$ (bottom) as a function of drive power extracted from Ramsey interferometry without cancellation (blue) and with cancellation (red). Error bars are statistical. Solid lines: Numerical simulations of the measurement-induced dephasing and ac-Stark shift. The simulation which accounts exactly for the $\cos\hat\varphi_\mathrm{tr}$ potential of the Josephson junction is performed using the bare parameters $\omega_r/2\pi = 7.66$ GHz, $E_J/h = 16.83$ GHz, $E_C/h = 199.7$ MHz, $g/2\pi = 140.6$ MHz, and $\kappa/2\pi = 10.1$ MHz resulting in dressed parameters that match the experimentally measured ones. The cavity drive frequency is $\omega_1 /2\pi = \SI{7.6648}{GHz}$.}
\label{fig: acStark-dephasing}
\end{figure}

As a second example, consider the typical situation of a dispersive qubit readout where the frequency of the first drive is resonant with the cavity and the qubit-cavity detuning is large compared to $g'$ such that the system is in the dispersive regime \cite{Haroche2006,Blais2021}. In the absence of the cancellation tone, the cavity drive results in ac-Stark shift and broadening of the qubit~\cite{Schuster2005,Gambetta2006}. This is illustrated in the top panel of \cref{fig: acStark-dephasing}a which shows the numerically computed magnitude of the qubit's absorption spectrum $S(\omega)$. For simplicity, this is obtained in the two-level approximation for the transmon where the absorption spectrum takes the form $S(\omega) = (1/2\pi) \int_{-\infty}^{+\infty} e^{i\omega t}\mean{\hat \sigma_-(t) \hat \sigma_+(0)}_s$ with the subscript $s$ indicating that the average is taken in steady-state. The different colored lines correspond to different cavity drive amplitudes $\varepsilon_1$. The vertical gray line indicates the bare qubit frequency. In contrast, the bottom panel of \cref{fig: acStark-dephasing}a shows the same quantity obtained with the cancellation tone present. In this case, the results for the different drive amplitudes collapse on each other. Because cloaking does not affect vacuum fluctuations, the Lamb shift (i.e.~the offset from the bare qubit frequency) remains unchanged at all drive amplitudes. Moreover, the independence of the qubit linewidth on drive power indicates both the absence of measurement-induced dephasing and that Purcell decay remains at its zero-photon value~\cite{Boissonneault2010,Sete2014} under cloaking. 

This prediction is experimentally tested on the device shown in \cref{fig: cartoon}c. It consists of a transmon qubit capacitively coupled to a $\lambda/4$ microwave coplanar waveguide cavity which is driven through port 1 ($\mathcal{E}_1$) via a Purcell filter acting as a bandpass filter at the cavity frequency. The cancellation drive ($\mathcal{E}_2$) is applied to port 2, which is weakly capacitively coupled to the transmon. The readout frequency is $\omega_1 /2\pi = \SI{7.6648}{GHz}$ and cavity decay rate $\kappa/2\pi = \SI{10.1}{MHz}$. The transmon qubit is {capacitively coupled to the cavity, with $\chi/2\pi = \SI{-2.54}{MHz}$ the full dispersive shift~\cite{Blais2021}}. The qubit has a dressed frequency $\tilde \omega_q/2\pi = \SI{4.96216}{GHz}$ and coherence times $T_1=\SI{25}{\mu s}$ and $T_2 = \SI{7.5}{\mu s}$.

Using Ramsey interferometry, it is possible to determine the decoherence rate and the qubit frequency, and thus extract the increase in dephasing rate $\delta\Gamma$ and the ac-Stark shift $\delta\omega$ as a function of drive strength. The blue dots in \cref{fig: acStark-dephasing}b are obtained in the absence of cancellation drive and show the expected linear increase with drive power~\cite{Schuster2005,Gambetta2006}. The full lines are obtained from numerical integration of the system's master equation including cavity loss and are used to calibrate the attenuation factor between the drive power at room temperature and $\epsilon_1^2$ for all experimental plots. The red dots correspond to the measured ac-Stark shift and increase in dephasing rate in the presence of the cancellation tone. As expected from the above discussion, with the properly tailored cancellation drive, they are both suppressed at all cavity drive powers. The cancellation drive follows the analytical expression \cref{eq: E2} for $\mathcal{E}_2(t)$ but the attenuation and electrical delay of the line driving port 2 needs to be taken into account in order to relate the complex drive amplitude at room temperature $\mathcal{E}^\mathrm{room}_2(t)$ to the driving strength $\mathcal{E}_2(t)=\mu \mathrm{Re}[e^{i\phi}\mathcal{E}^\mathrm{room}_2(t)]$. The prefactor relating the two is found experimentally by varying the phase $\phi$ and amplitude $\mu$ to minimize $|\delta\Gamma+i\delta\omega|$ (see Supplementary Note 2~\cite{sm}). This technique thus provides a tool to calibrate the attenuation from room temperature to the qubit port.

\begin{figure}[t!]
\includegraphics{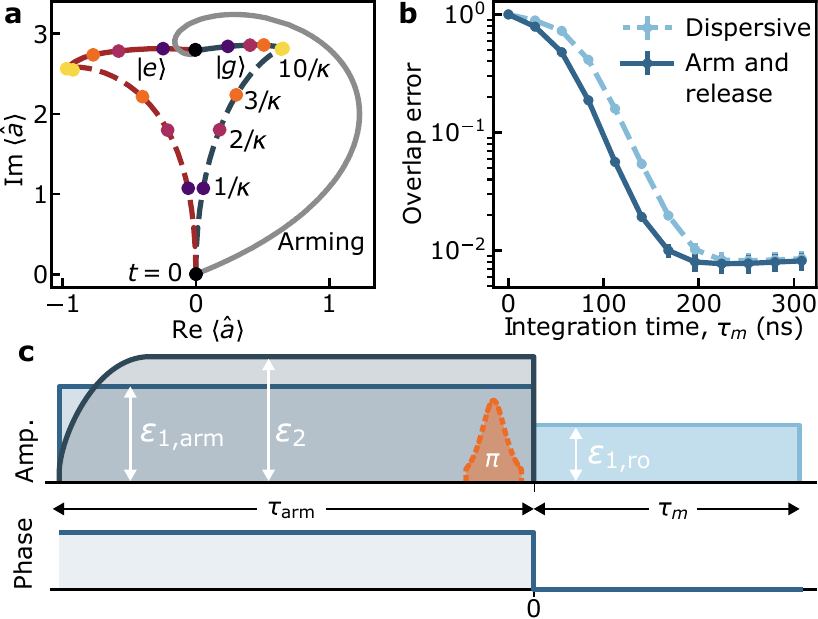}
\caption{\captiontitle{Arm and release qubit readout.} \textbf{a}, Full-cosine numerical simulation of the phase-space trajectories of the cavity field in standard dispersive readout, i.e., square pulse, (dashed line) and the arm and release readout approach (full lines) with an arm-step coherent state $\alpha=2.8i$. The path of the arming step is here chosen to reproduce the experiment of panel \textbf{b}, but it can be tailored at will. The coloured dots indicate different times in the evolution. In both readout approaches, the qubit is prepared in the ground state and a $\pi$ pulse is applied (red lines) or not (blue lines) before the readout step. Simulation parameters are the same as in \cref{fig: acStark-dephasing}b. \textbf{b}, Experimental qubit measurement error obtained from the overlap between the distributions of the accumulated heterodyne signal over $10^6$ repetitions of the experiment. Full line: arm-and-release approach. Dashed line: standard dispersive measurement. \textbf{c}, Arm-and-release pulse sequence used to obtain panels \textbf{a} and \textbf{b}. The arm phase ($t<0$) is absent in the case of dispersive readout.
}
\label{fig: Faster readout}
\end{figure}

The above results suggest a strategy to speed up the dispersive qubit readout. Several approaches have been explored to improve readout fidelity by speeding it up~\cite{Jeffrey2014, McClure2016, Bultink2016, Touzard2019, Ikonen2019, Peronnin2020, Didier2015}. Here, we propose to use a two-step `arm and release' approach. With the cavity in the vacuum state, the arming step consists in driving the cavity in the presence of the cancelling tone. Because the qubit is uncoupled from the cavity's classical field, this causes a displacement of the cavity field by $\alpha_t$ that is \emph{independent} of the qubit-state. Thus during the pre-arming, the cavity can be stabilized in any coherent state without affecting the qubit. The release step can start at any time after the desired measurement photon population is reached: The cancellation tone is turned off, at which point the qubit couples to the cavity field resulting in a qubit-state \emph{dependent} rotation of the latter in phase space under the dispersive qubit-cavity coupling. Homodyne or heterodyne detection then completes the qubit readout in the usual way~\cite{Blais2021}.

Crucially, upon release the two coherent states $|\alpha_{g,e}\rangle$ corresponding to the qubit ground and excited states separate as under longitudinal coupling at short times. Indeed, as illustrated in \cref{fig: Faster readout}a, after the arming step these coherent states move away from each other in the phase space of the cavity mode (full lines), thus maximizing the qubit measurement rate $\kappa|\alpha_g-\alpha_e|^2/2$~\cite{Gambetta2006,Didier2015}. This is reminiscent of previous experiments~\cite{Ikonen2019,Touzard2019} where longitudinal-like separation is obtained with an initially empty cavity using an approach based on the dispersive approximation. In contrast, we emphasize that qubit cloaking can operate at arbitrary drive strength and qubit-cavity detuning. \Cref{fig: Faster readout}a also shows the evolution in phase space of the amplitude of the readout cavity for the usual dispersive readout (dashed line), where there is initially poor separation between the qubit-state-dependent coherent states. Similarly to longitudinal readout~\cite{Didier2015}, for the same steady-state photon population (see yellow dots corresponding to $t = 10/\kappa$), the state separation is significantly larger at short times in the arm-and-release approach than in the usual dispersive readout (see the coloured dots corresponding to times $t=(1,2,3)/\kappa$).

The reduction in measurement time provided by the arm-and-release approach is illustrated in \cref{fig: Faster readout}b which shows the measurement error versus integration time for the arm-and-release approach (full line) and the standard dispersive readout (dashed line) obtained using the device of \cref{fig: cartoon}c. In these experiments, the qubit is first prepared in the ground state $\ket{g}$ using measurement-based feedback with the usual dispersive readout. As illustrated in \cref{fig: Faster readout}c, to obtain the full line in panel b the cavity is then pre-armed with a  drive of amplitude $\varepsilon_{1,\textrm{arm}}/2\pi = \SI{95.5}{MHz}$ (full blue line) and a cancellation drive $\varepsilon_2$ (dashed dark blue line) for a time $\tau_\mathrm{arm}=\SI{1}{\mu s}$. At the time labelled $t=0$, the signal is then integrated for a time $\tau_m$ with a cavity drive $\varepsilon_{1,\mathrm{ro}}/2\pi = \SI{63.7}{MHz}$. The arming cavity drive and the cancellation tone are omitted to obtain the dashed line in panel b. A $\pi$ pulse of width $\SI{25}{ns}$ (orange dashed line) is applied or not at the end of the arming step to prepare the state $\ket{e}$ or $\ket{g}$. Interestingly and as explained below, this can be done with high fidelity despite the cavity photon population. Finally, the measurement error is obtained by computing the overlap between the distribution of the accumulated heterodyne signal over $N = 10^6$ repetitions of the experiment where the qubit is prepared either in $\ket{g}$ or $\ket{e}$ (see Supplementary Note 2~\cite{sm}). For a 196 ns integration time, we obtain an average fidelity $\mathcal{F} = 1 - [P(g|e)+P(e|g)]/2 = \SI{99.35 \pm 0.14}{\%}$ where $P(g|e)=\SI{1.07\pm 0.14}{\%}$ is the error probability to measure state $g$ when state $e$ was prepared, while $P(e|g)=\SI{0.23\pm 0.14}{\%}$ is the error probability to measure state $e$ when state $g$ was prepared. The finite number of repetitions $N$ results in the uncertainty $\pm\SI{0.14}{\%}$. The $\SI{0.65}{\%}$ average error is mostly explained by wrong preparation of the ground state before the arming step ($\sim 0.2\%$), imperfect $\pi$ pulse ($\sim 0.08\%$), relaxation during measurement ($\sim 0.2\%$), and finite Gaussian separation ($\sim 0.13\%$).

In optimizing readout fidelity, it is important to account that the cavity responds at different frequencies in the absence or presence of the cancellation tone. Without cloaking, the qubit-state-dependent steady-states coherent state amplitude is $\alpha_{i}^\text{s} = (\varepsilon_1 e^{-i\phi_1}/2)(\tilde \omega_r + \chi_i - \omega_1 - i\kappa/2)$, where $\tilde \omega_r$ is the dressed cavity frequency and $\chi_i$ is the dispersive qubit-cavity coupling for qubit state $i=\{g,e\}$~\cite{Blais2021}. On the other hand, with cloaking the cavity responds as if there was no qubit with the steady-state value $\alpha^\text{s} = (\varepsilon_1 e^{-i\phi_1}/2)/(\omega_r-\omega_1-i\kappa/2)$, where now $\omega_r$ is the bare cavity frequency. Depending on the application, such as optimizing the longitudinal-like nature of the readout, an optimal coherent state can be prepared by adequately choosing $\varepsilon_1$ and $\phi_1$ during the arming phase, and changing these quantities as desired during the release phase; see \cref{fig: Faster readout}c for the pulse sequence and phase used here for readout. The arming phase-space path shown in \cref{fig: Faster readout}a is chosen to replicate our experiment, but it can be tailored. For example, a straight path is obtained by arming the cavity with a drive frequency $\omega_1 = \omega_r$.

\begin{figure}[t!]
\includegraphics{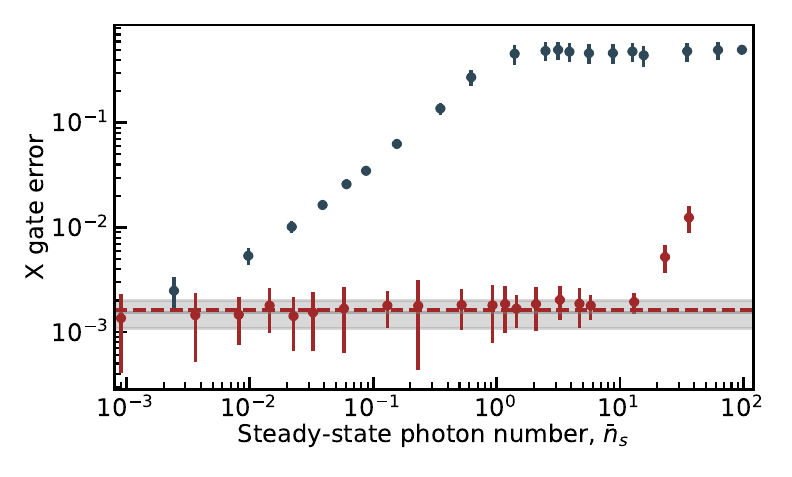}
\caption{\captiontitle{Gate error.} Measured error (dots) on qubit X gate (randomized benchmarking) as a function of the steady-state cavity photon number $\bar n_\mathrm{s}$ when the qubit is cloaked (red) or not (blue).
The control drive on the qubit is optimized in the absence of cavity drive and cancellation tone to maximize the gate fidelity, and kept identical for all further measurements. The corresponding measured average X gate error and its error bar at zero cavity drive amplitude are represented as a horizontal gray line and gray area. The dashed red line is the average gate error under cloaking obtained from numerical simulations including the Purcell filter (see Supplementary Note 1~\cite{sm}). Error bars account for statistical uncertainty.}
\label{fig: Gates}
\end{figure}

As an additional benefit, since the cloaked qubit does not undergo ac-Stark shift or measurement-induced dephasing (see \cref{fig: acStark-dephasing}), it is possible to apply qubit gates during the cavity arming step despite the presence of measurement photons in the cavity. As a result, the time needed to fill the cavity does not factor in the measurement time. To test this idea, we use randomized benchmarking~\cite{Dankert2009, Magesan2011} on the device of \cref{fig: cartoon}c, and extract the gate error for a qubit $\pi$-pulse (X gate) performed in the presence of a readout cavity drive, and in the absence or presence of the cancellation tone, see \cref{fig: Gates}. As expected, in the absence of cancellation (blue dots) the gate error increases rapidly with the cavity drive amplitude and saturates at a gate error of 0.5, corresponding to the largest gate error which can be reported by randomized benchmarking~\cite{Knill2008}. In the presence of cancellation (red dots), the gate error remains at its coherence-limited value (full gray line) despite the presence of measurement photons in the readout cavity. The red dashed line corresponds to the result of numerical simulations accounting for the Purcell filter of the gate under cloaking, see Supplementary Note 1~\cite{sm}. Importantly, the qubit control drive is identical for all data points shown in \cref{fig: Gates}. For large drive amplitudes, the X gate error under cloaking starts to increase (last three points in \cref{fig: Gates}). This is explained by imperfect experimental calibration of the cancellation tone which can be affected by low-frequency drifts.

\section*{Discussion}

Qubit cloaking can readily be implemented in current circuit QED experiments. This approach is valid irrespective of the driving frequency and waveshape, and applies to arbitrary qubit-cavity detuning. In the resonant regime, cloaking can be used to prepare non-classical states of the cavity~\cite{Uria2020, Long2022}. In the dispersive regime, cloaking can be used to speed-up qubit readout~\cite{Munoz2023} something which, for example, can be used to shorten quantum error correction cycles in circuit QED-based devices~\cite{ai2021exponential, Krinner2022, Zhao2022}. At the heart of this acceleration is the fact that logical gates can be applied to the qubit while the cavity is armed for readout. Similar ideas can be exploited to speed up two-qubit gates that are assisted by a cavity drive, such as the resonator-induced phase gate~\cite{Paik2016} or to reduce measurement cross-talk errors in multiplexed readout~\cite{Heinsoo2018}. Moreover, the possibility to cancel large ac-Stark shifts can be leveraged to avoid unwanted phase accumulations, such as in the preparation of bosonic codes states~\cite{Eickbusch2022}. We expect that qubit cloaking will become a useful element in the toolbox of circuit QED, extending beyond the transmon qubit~\cite{Gyenis2021}. Furthermore, it is applicable to other cavity QED setups where a quantum system linearly couples to a cavity~\cite{Mi2017, Stockklauser2017, Samkharadze2018, Keyser2020, Manenti2017, Ye1999}.

\section*{Data Availability Statement}
The data generated in this study have been deposited in the Figshare database~\cite{Repository-data}.

\bibliography{references}

\section*{Acknowledgements}
The authors are grateful to \'Elie Genois for helpful discussions and to Lincoln Labs and IARPA from providing the TWPA.
\begin{itemize}
\item NSERC -- A. Moulinas and A. Blais\\
\item Canada First Research Excellence Fund -- C. Lledó, A. Moulinas, J. Cohen, R. Shillito, and A. Blais\\
\item U.S. Army Research Office Grant No.~W911NF-18-1-0411 -- J. Cohen, R. Shillito, and A. Blais\\
\item Minist\`ere de l’\'Economie et de l’Innovation du Qu\'ebec -- C. Lledó and A. Blais\\
\item France 2030 grant ANR-22-PETQ-0006 R. Dassonneville, A. Bienfait, and B. Huard \\
\item ANR grant ANR-21-CE47-0007 -- R. Dassonneville, A. Bienfait, and B. Huard
\end{itemize}

\section*{Author Contributions Statement}
C. Lledó and R. Dassonneville contributed equally.

C. Lledó and A. Moulinas performed the calculations.\\
R. Dassonneville performed the measurements.\\
R. Shillito provided codes.\\
C. Lledó, R. Dassonneville, A. Moulinas, J. Cohen, A. Bienfait, B. Huard, and A. Blais analyzed and discussed the results.\\
C. Lledó, R. Dassonneville, and A. Blais wrote the manuscript with comments from all the authors.\\
A. Bienfait, B. Huard, and A. Blais supervised the work.

\section*{Competing interests}
B. H. is an equity shareholder of the Alice\&Bob company. The remaining authors declare no competing interests.

\newpage

\end{document}


\author{Crist\'obal Lled\'o$^*$}
\affiliation{Institut Quantique and D\'epartement de Physique, Universit\'e de Sherbrooke, Sherbrooke J1K 2R1 QC, Canada}
\author{R\'emy Dassonneville$^*$}
\affiliation{Ecole Normale Sup\'erieure de Lyon, CNRS, Laboratoire de Physique, F-69342 Lyon, France}
\author{Adrien Moulinas}
\affiliation{Institut Quantique and D\'epartement de Physique, Universit\'e de Sherbrooke, Sherbrooke J1K 2R1 QC, Canada}
\author{Joachim Cohen}
\affiliation{Institut Quantique and D\'epartement de Physique, Universit\'e de Sherbrooke, Sherbrooke J1K 2R1 QC, Canada}
\author{Ross Shillito}
\affiliation{Institut Quantique and D\'epartement de Physique, Universit\'e de Sherbrooke, Sherbrooke J1K 2R1 QC, Canada}
\author{Audrey Bienfait}
\affiliation{Ecole Normale Sup\'erieure de Lyon,  CNRS, Laboratoire de Physique, F-69342 Lyon, France}
\author{Benjamin Huard}
\affiliation{Ecole Normale Sup\'erieure de Lyon,  CNRS, Laboratoire de Physique, F-69342 Lyon, France}
\author{Alexandre Blais}
\affiliation{Institut Quantique and D\'epartement de Physique, Universit\'e de Sherbrooke, Sherbrooke J1K 2R1 QC, Canada}
\affiliation{Canadian Institute for Advanced Research, Toronto, M5G1M1 Ontario, Canada}

{\let\thefootnote\relax\footnote{{${}^*$ These authors contributed equally to this work.}}}

\title{Supplementary Information for Cloaking a qubit in a cavity}

\date{\today}

\maketitle

\section*{Supplementary Note 1 -- Theory}

\subsection{Qubit cloaking: detailed derivation}
\label{SM section: detailed derivation of the cancellation method}

We start from the Lindblad equation 
\begin{equation}\label{eq:SimpleME}
\partial_t \hat \rho = -\frac{i}{\hbar}[\hat H, \hat \rho] + \kappa\mathcal D[\hat a]\hat \rho,    
\end{equation}
where $\hat H = \hat H_0 + \hat H_1 + \hat H_2$, and $\mathcal D[\hat a]\hat \rho = \hat a \hat \rho \hat a^\dag - (1/2)\{\hat a^\dag \hat a, \hat \rho\}$ is the Lindblad dissipator  describing photon decay~\cite{Breuer_book}. We move to the displaced frame $\hat \rho_D(t) = \hat D^\dag (\alpha_t) \hat \rho(t) \hat D(\alpha_t)$ using the displacement operator $\hat D(\alpha_t) = \exp(\alpha_t\hat a^\dag - \alpha_t^* \hat a)$. In this frame, the evolution of the system state is governed by the equation
\begin{equation}\label{equ SM: displaced frame Lindblad equation}
\begin{split}
    \partial_t \hat \rho_D =& -i[\hat H_\mathrm{tr}/\hbar + \omega_r \hat a^\dag \hat a + i g\hat n_\mathrm{tr}(\hat a^\dag - \hat a), \hat \rho_D]  \\
    &+ \kappa \mathcal D[\hat a]\hat \rho_D\\
    & -i[\mathcal{E}_2(t)\hat n_\mathrm{tr} - ig(\alpha_t- \alpha_t^*) \hat n_\mathrm{tr}, \hat \rho_D] \\
    &+[\hat a, \hat \rho_D](-\dot \alpha_t^* + (i \omega_r - \kappa/2)\alpha_t^* + \mathcal E_1(t)) \\
    &+[\hat a^\dag, \hat \rho_D](-\dot \alpha_t- (i\omega_r + \kappa/2)\alpha_t+ \mathcal E_1(t)),
\end{split}
\end{equation}
where we have introduced the transmon Hamiltonian $\hat H_\mathrm{tr} = 4E_C \hat n_\mathrm{tr}^2 - E_J \cos(\hat \varphi_\mathrm{tr})$. The form of the time-dependent complex amplitude $\alpha_t$ is chosen such as to cancel the cavity drive. For this, the last two lines in \cref{equ SM: displaced frame Lindblad equation} must vanish and we thus enforce
\begin{equation} \label{equ: equation for alpha(t)}
\dot \alpha_t = -(i\omega_r + \kappa/2)\alpha_t + \mathcal E_1(t),
\end{equation}
or, equivalently,
\begin{equation} \label{equ SM: full alpha(t) solution}
\begin{split}
    \alpha_t =& \alpha_0 e^{-(i\omega_r + \kappa/2)t} \\
    &+ \int\limits_{0}^{t} d\tau\, e^{-(i\omega_r + \kappa/2)(t-\tau)}\mathcal E_1(\tau),
\end{split}
\end{equation}
as mentioned in the discussion following Eq.(3) in the main text. In this frame, the cavity drive is then effectively passed to the qubit, see the third line of \cref{equ SM: displaced frame Lindblad equation}. The latter is cancelled if we choose
\begin{equation} \label{equ SM: cancelling tone}
    \mathcal{E}_2(t) = -ig(\alpha_t^* - \alpha_t) = -2 g \text{Im}[\alpha_t].
\end{equation}
The initial condition $\alpha_0$ can be set to zero; it has no consequence on cloaking and, in this way, $\mathcal E_2(t)$ starts from zero at $t=0$.

Also worth noting is that accounting for thermal incoherent excitations in the cavity---replacing $\kappa \mathcal{D}[\hat a]\hat \rho$ by $(\bar{n}_\text{th}+1)\kappa \mathcal{D}[\hat a]\hat \rho + \bar{n}_\text{th}\kappa \mathcal D[\hat a^\dag]\hat \rho$ in \cref{eq:SimpleME}, where $\bar n_\text{th}$ is the thermal population---, qubit decay and dephasing, or incorporating the multimode nature of the cavity, does not change the effect or form of the cancellation drive. We come back to incoherent cavity excitation in \cref{sec: Potential limitations}.

For a constant cavity drive amplitude $\varepsilon_1$ turned on at $t=0$, the cancellation tone takes the form
\begin{equation} \label{equ SM: cancellation tone for constant cavity drive}
\begin{split}
\mathcal{E}_2(t) =& g \varepsilon_1 \left[ \frac{\cos(\omega_1 t + \phi_{1,+}) - \cos(\omega_r t - \phi_+)e^{-\kappa t/2}}{\sqrt{\omega_+^2 + (\kappa/2)^2}} \right. \\
&\quad -\left. \frac{\cos(\omega_1 t - \phi_{1,-}) - \cos(\omega_r t - \phi_-)e^{-\kappa t/2}}{\sqrt{\omega_-^2 + (\kappa/2)^2}}  \right],
\end{split}
\end{equation}
where $\omega_\pm = \omega_r \pm \omega_1$, $\phi_\pm = \arctan(-2\omega_\pm/\kappa)$, and $\phi_{1,\pm} = \phi_1 + \phi_{\pm}$. 
With $\omega_r \gg \kappa$ and taking $\omega_1 \sim \omega_r$, the leading terms are the ones in the second line of \cref{equ SM: cancellation tone for constant cavity drive} and the cancellation tone can be approximated by
\begin{equation} \label{equ SM: approximate cancellation tone for a constant cavity drive}
\begin{split}
\mathcal{E}_2(t) \approx  A_- &\big[\cos(\omega_1 t - \phi_{1,-}) - \cos(\omega_r t - \phi_-)e^{-\kappa t/2} \big],
\end{split}
\end{equation}
with $A_- = -g \varepsilon_1/\sqrt{\omega_-^2 + (\kappa/2)^2}$. In experiments, a cancellation tone ansatz of the form \cref{equ SM: approximate cancellation tone for a constant cavity drive} can be used where $A_-$, $\phi_{1,-}$, and $\phi_-$ are unknown parameters that can be optimized in a  Ramsey-like experiment by minimizing the ac-Stark shift and measurement-induced dephasing, see Fig.~3 in the main text. Importantly, the use of the cancellation tone in \cref{equ SM: cancellation tone for constant cavity drive} or~(\ref{equ SM: approximate cancellation tone for a constant cavity drive}) demands knowledge of the bare cavity frequency $\omega_r$. An incorrect estimate for $\omega_r$ will lead to imperfect cancellation, but only for transient times before the terms oscillating at $\omega_r$ in \cref{equ SM: cancellation tone for constant cavity drive} or~(\ref{equ SM: approximate cancellation tone for a constant cavity drive}) become negligible due to the exponential decay $e^{-\kappa t/2}$.

In practice, there can be small crosstalks by which the drive on the cavity (qubit) also weakly drives the qubit (cavity). Formally, for a known $\mathcal E_1(t)$, the simple form of $\mathcal E_2(t) = -2g \mathrm{Im}[\alpha_t]$ which achieves the cancellation is replaced in this case by a linear Volterra equation of the second kind: $\mathcal E_2(t) = f(t) + \int_0^t d\tau K(t, \tau)\mathcal E_2(\tau)$ with a separable Kernel $K(t,\tau) = \sum_j g_j(t)h_j(\tau)$. This equation is guaranteed to have a unique solution. In practice, the experimental calibration of the cancellation tone's amplitude and phase deals with this potential crosstalk.

\subsection{Two-states and RWA approximations for the transmon}

In the two-level approximation for the transmon and using the rotating wave approximation (RWA), the Hamiltonian $\hat H_0 + \hat H_1$ becomes
\begin{equation}\label{eq:H01-TLS}
\begin{split}
\hat H_0 + \hat H_1 =& \frac{\hbar\omega_q}{2} \hat \sigma^z  + \hbar\omega_r \hat a^\dag \hat a + \hbar g'(\hat \sigma^+ \hat a + \hat \sigma^- \hat a^\dag) \\
&- \frac{\hbar\varepsilon_1(t)}{2} (\hat a^\dag e^{-i\omega_1 t} + \hat a e^{i\omega_1 t}).
\end{split}
\end{equation}
In terms of the transmon parameters~\cite{Koch2007}, the qubit frequency is given by $\hbar\omega_q \approx \sqrt{8 E_C E_J} - E_C$ and the coupling is $g' \approx (g/2) (E_J/2E_C)^{1/4}$. With these approximations, the cancellation tone on the qubit takes the form
\begin{equation}\label{eq:H2-TLS}
\hat H_2 = \hbar g'(\text{Re}[\alpha_t] \hat \sigma^x - \text{Im}[\alpha_t] \hat \sigma^y),
\end{equation}
with
\begin{equation}
    \alpha_t = \frac{i}{2} \int\limits_{0}^{t} d\tau\, e^{-(i\omega_r + \kappa/2)(t-\tau)}\varepsilon_1(\tau) e^{-i\omega_1 \tau}.
\end{equation}
To obtain the results of Fig.~2 and Fig.~3a of the main text, we numerically integrate the master equation of \cref{eq:SimpleME} with the Hamiltonians of \cref{eq:H01-TLS} and \cref{eq:H2-TLS}.

\subsection{Cloaking of a general multilevel quantum system}

As is made clear by the above two examples, as long as the coupling Hamiltonian is linear in the cavity mode creation and annihilation operators ($\hat a$ and $\hat a^\dag$), the approach proposed here is applicable to a wide range of quantum systems where two driving ports couple asymmetrically to the cavity and the quantum system. Indeed, we can generalize our cloaking method for any multilevel quantum system with Hamiltonian $\hat H_q$, coupled to a cavity with $\hat H_g = g \hat O_q (e^{-i\phi} \hat a^\dag + e^{i\phi} \hat a)$ via any quadrature (i.e. arbitrary phase $\phi$), where $\hat O_q$ is any operator on the multilevel system. 

For this general case, the cancellation term $\hat H_2(t)$ in the Lindblad equation $\partial_t \hat \rho = -\frac{i}{\hbar}[\hat H, \hat \rho] + \kappa \mathcal D[\hat a](\hat \rho)$, where
\begin{equation}
\begin{split}
\hat H =& \hat H_q + \hat H_g + \hbar\omega_r \hat a^\dag \hat a \\
&+ i\hbar\mathcal E_1(t) (\hat a^\dag - \hat a) + \hat H_2(t),
\end{split}
\end{equation}
is given by
\begin{equation} \label{equ SM: general cancellation tone}
\hat H_2(t) = -g(\alpha_t e^{i\phi} + \alpha_t^* e^{-i\phi}) \hat O_q,
\end{equation}
with $\alpha_t$ as in \cref{equ SM: full alpha(t) solution}.

\subsection{Inclusion  of the Purcell filter}

Considering a general multilevel system with Hamiltonian $\hat H_q$ as in the previous subsection, we now account for the presence of a Purcell filter cavity coupled to the readout cavity, such that the total Hamiltonian reads
\begin{equation} \label{equ SM: Hamiltonian with Purcell filter}
\begin{split}
\hat H(t)=& \hat H_q + \hat H_g + \hbar\omega_r \hat a^\dag \hat a + \hat H_2(t) \\
&+ \hbar\omega_f \hat f^\dag \hat f + \hbar J (\hat a^\dag + \hat a)(\hat f^\dag + \hat f) \\
&+ i\hbar\varepsilon_1(t)\sin(\omega_1 t + \phi_1) (\hat f^\dag - \hat f),
\end{split}
\end{equation}
where $\hat f^{(\dag)}$ is the annihilation (creation) operator of the Purcell mode of frequency $\omega_f$, and now the drive is on the Purcell cavity instead of the readout cavity.

In the presence of photon loss at the Purcell filter cavity, the master equation takes the form 
\begin{equation} \label{equ SM: Lindblad equation with Purcell filter}
\partial_t \hat \rho = - \frac{i}{\hbar}[\hat H(t), \hat \rho] + \kappa_f \mathcal D[\hat f]\hat \rho.
\end{equation}
Following the above approach, we now move to a displaced frame $\hat \rho_D(t) = \hat D_r^\dag(\alpha_t) \hat D_f^\dag(\beta_t) \hat \rho(t) \hat D_r(\alpha_t) \hat D_f(\beta_t)$, where the subscripts $r$ and $f$ indicate that the displacement operator acts on the Hilbert space of the readout cavity mode $\hat a$ or the Purcell filter cavity mode $\hat f$, respectively. Moreover, we choose $\alpha_t$ and $\beta_t$ such that they follow the coupled equations of motion
\begin{equation} \label{equ SM: coupled semiclassical equations readout + Purcell}
\begin{split}
\partial_t \alpha_t=& -i\omega_r \alpha_t - i J(\beta_t^* + \beta_t) \\
\partial_t \beta_t =& -(i\omega_f + \kappa_f/2) \beta_t - iJ(\alpha_t^* + \alpha_t), \\
&+ \varepsilon_1(t) \sin(\omega_1 t + \phi_1).
\end{split}
\end{equation}
With this choice, the drive on the Purcell cavity is cancelled and effectively appears on the qubit. This effective drive on the qubit in the displaced frame is cancelled by the Hamiltonian $\hat H_2(t)$ in \cref{equ SM: general cancellation tone}, with $\alpha_t$ now the solution to \cref{equ SM: coupled semiclassical equations readout + Purcell}.

\subsection{Gates and imperfect cancellation}

Here we report on the simulation of logical gates on the transmon in the presence of a cavity drive. While the gate fidelity can be maintained constant for any cavity drive amplitude if the exact cancellation tone is used, here we explore the deviations that could arise if the cancellation tone is inexact.

As an example, consider the approximate cancellation in \cref{equ SM: approximate cancellation tone for a constant cavity drive} which is very precise for $\omega_1 \sim \omega_r \gg \kappa$. Among other kinds of errors, here we consider that resulting from using the simpler ansatz
\begin{equation} \label{equ SM: anzat cancellation tone}
    \mathcal{E}_2^\text{ansatz}(t) = A \cos(\omega_1 t - \phi)(1-e^{-\kappa t/2})
\end{equation}
and study how the gate fidelity deviates from the ideal result.

We consider the Lindblad equation
\begin{equation} \label{equ SM: Lindblad eq with gate}
\partial_t \hat \rho = -\frac{i}{\hbar}[\hat H + \hat H_\text{gate}, \hat \rho] + \kappa \mathcal D[\hat a]\hat \rho,
\end{equation}
where $\hat H$ includes the transmon qubit and cavity Hamiltonian $\hat H_0$, the cavity drive $\hat H_1$, and the cancellation $\hat H_2$ (if present). The gate consists in a $\pi$-rotation of $t_\text{g} = 25$ ns duration. Using DRAG to eliminate leakage~\cite{Motzoi2009}, the gate Hamiltonian reads
\begin{equation} \label{equ SM: gate Hamiltonian}
\begin{split}
\hat H_\text{gate} =& \hbar\varepsilon_\text{g} \left\{ \frac{1}{\cosh[f(t)]} \sin(\tilde \omega_q t + \phi_g) \right. \\
& + \left. d_g \frac{\sinh[f(t)]}{\cosh^2[f(t)]} \cos(\tilde \omega_q t + \phi_g) \right\} \hat n_\mathrm{tr}
\end{split}
\end{equation}
where $f(t) = \sqrt{\frac{\pi}{2}} (t-t_\text{g}/2)/(t_\text{g}/4)$, $\tilde \omega_q$ the transmon $g-e$ transition frequency dressed by the cavity, and \{$\varepsilon_g, d_g, \phi_g\}$ the set of parameters which are optimized to realize high-fidelity operations. This optimization is performed only once in the absence of cavity drive and cancellation tone. We note that this waveform for the $\pi$ pulse is the same as the one we use in our experiment.

Numerically, to characterize the gate fidelity, we use~\cite{Bowdrey2002}
\begin{equation} \label{equ SM: average gate fidelity no drives}
\mathcal F_\text{no drives} = \frac{1}{6}\sum_{i} \Tr[\hat U \hat \rho_i \hat U^\dag \mathcal E_{t_g}(\hat \rho_i)],
\end{equation}
which averages over the six cardinal states ($i=\pm x, \pm y, \pm z$) of the Bloch sphere in the dressed computational subspace $\{\ket{\overline{g,0}}, \ket{\overline{e, 0}} \}$, corresponding to the qubit logical states dressed by the cavity~\cite{Blais2021}. The operator $\hat U$ is the $X$ Pauli operator in this subspace and $\mathcal E_{t_g}$ is the dynamical map which evolves the state according to the master equation of \cref{equ SM: Lindblad eq with gate}. 

Using $E_J/h = 84.28 E_C/h = 16.826$ GHz, $\omega_r/2\pi = 7.657$ GHz, $g/2\pi = 140.6$ MHz, $\kappa/2\pi = 10.1$ MHz, and accounting for three transmon energy levels, we obtain the optimal values $\varepsilon_g/2\pi \approx 28.51$ MHz, $d_g \approx 0.09$, and $\phi_g/2\pi \approx -0.027$, producing a gate error $E=1-\mathcal F_\text{no drives} \approx 0.003$ in the absence of cavity and cancellation tones. This value corresponds to the coherence limit ($\sim 1-e^{-\gamma_\kappa t_g/2}$) set by the Purcell decay rate $\gamma_\kappa$~\cite{Blais2021}.

In the presence of the cavity drive and in the displaced frame introduced in the main text and in \cref{equ SM: displaced frame Lindblad equation}, the gate realizes a high-fidelity X $\pi$-rotation within the subspace $\{\ket{\overline{g,0}}, \ket{\overline{e, 0}} \}$ as long as the qubit is cloaked. Back in the laboratory frame, the dynamical evolution starting with the qubit in an arbitrary state and the cavity in the vacuum state, turning on the cavity drive and the cancellation tone, and performing the logical gate corresponds to the transformation
\begin{equation}
\psi_g \ket{\overline{g,0}} + \psi_e \ket{\overline{e,0}} \to \hat D(\alpha_{t_g}) (\psi_g \ket{\overline{e,0}} + \psi_e \ket{\overline{g,0}}).
\end{equation}
To account for the displacement operator, the expression for the gate fidelity in \cref{equ SM: average gate fidelity no drives} in the presence of drive and cancellation is modified to
\begin{equation} \label{equ SM: average gate fidelity}
\mathcal F = \frac{1}{6}\sum_i \Tr[\hat \sigma_i \mathcal E_{t_g}(\hat \rho_i)],
\end{equation}
where
\begin{equation}
\hat \sigma_i = \hat D(\alpha_{t_g}) \hat U \hat \rho_i \hat U^\dag \hat D^\dag(\alpha_{t_g}).
\end{equation}
Note that for $\varepsilon_1=0$ (and thus $\mathcal{E}_2=0$),  $\alpha_{t_g}=0$ and this expression corresponds to \cref{equ SM: average gate fidelity no drives}.

\begin{figure}[t!]
\includegraphics{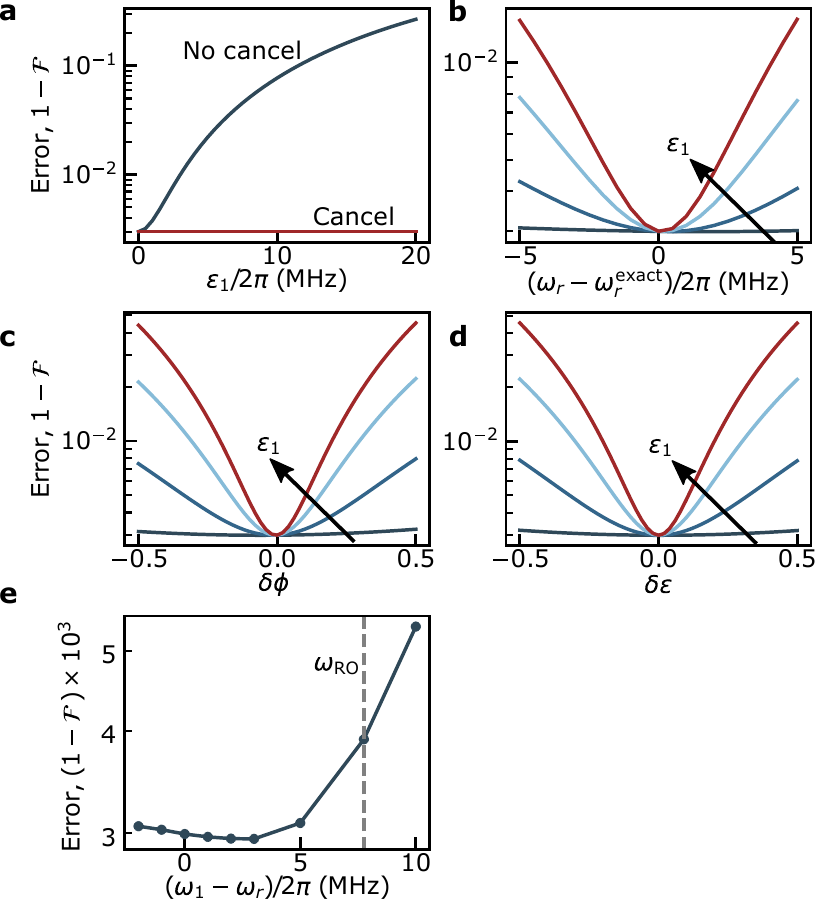}
\caption{\captiontitle{Numerical $X$-gate average error.} \textbf{a}, Error as a function of cavity drive amplitude without cancellation (blue line) and with exact cancellation (red line). \textbf{b}, Error when the cancellation tone in \cref{equ SM: approximate cancellation tone for a constant cavity drive} is used, but with a hypothetical imprecise estimate of the cavity frequency value, $\omega_r\neq \omega_r^\text{exact}$. Different colors correspond to $\varepsilon_1/2\pi =$ 1, 5, 10, and 15 MHz. The arrow indicates increasing values of $\varepsilon_1$. \textbf{c,d}, Error when there is, respectively, a phase ($\delta\phi$) or amplitude ($\delta \varepsilon$) relative offset (see text). Same $\varepsilon_1$ values as in \textbf{b}. In all \textbf{a,b,c,d}, the cavity is driven at the readout frequency, $\omega_1 = \omega_\text{RO}= \tilde \omega_r + (\chi_g + \chi_e)/2$. \textbf{e}, The simpler cancellation tone ansatz in \cref{equ SM: anzat cancellation tone} is used, which consist of a single frequency $\omega_1$. The drive amplitude is $\varepsilon_1/2\pi = \SI{15}{MHz}$. The vertical dashed line indicates the value of the readout frequency, $\omega_\mathrm{RO}$. The value of all parameters are indicated in the text.}
\label{fig SM: theory gates}
\end{figure}

In \cref{fig SM: theory gates}a we show the average gate error obtained from the above expression and integration of the master equation of \cref{equ SM: Lindblad eq with gate} as a function of $\varepsilon_1$ when the cancellation tone is off (blue line) and with the exact cancellation tone (red line). In the latter case and as expected from the discussion in the main text, the gate error is constant at $\sim 0.3\%$ for all drive amplitudes but rapidly increases with drive amplitude in the absence of cancellation. In \cref{fig SM: theory gates}b the gate error is plotted considering the approximate cancellation tone in \cref{equ SM: approximate cancellation tone for a constant cavity drive}, but in the hypothetical case of an imprecise estimate of the value of $\omega_r~(\neq \omega_r^\text{exact})$, highlighting the importance of correctly determining the bare cavity frequency to use in the cancellation tone. We reiterate that the imprecision leads to a slightly incomplete cancellation which is corrected after a short transient due to the exponential decay in time of the driving terms at the bare cavity frequency. When necessary, the bare frequency $\omega_r$ could be a parameter to optimize over in the experimental calibration of the cancellation tone. In \cref{fig SM: theory gates}c and d, we show the error when we use the cancellation tone in \cref{equ SM: cancellation tone for constant cavity drive} but with phase and amplitude relative offsets, respectively. In this case, we replace $\phi_\pm \to \phi_\pm(1+\delta \phi)$ or $\varepsilon_1 \to \varepsilon_1(1 + \delta \varepsilon)$ in the equation for $\mathcal E_2(t)$, taking $\phi_1=0$ for simplicity. In the case of phase offset, the main contribution to the deterioration of the gate comes from the offset in $\phi_-$, which enters in the leading terms (second line in \cref{equ SM: cancellation tone for constant cavity drive}). Finally, in \cref{fig SM: theory gates}e, we show the result of using the simpler anzat in \cref{equ SM: anzat cancellation tone} with a single frequency $\omega_1$, showing that when $\omega_1\approx \omega_r$ the gate maintains a high fidelity and deteriorates as the cancellation tone frequency shifts away.

\subsection{Gate simulation including Purcell cavity}

Here we give details about the simulations used to extract the calculated average $\pi$-rotation (X gate) error results shown in Fig.~5 in the main text, for the case of a cloaked qubit. Below we also show results for this error when the qubit is not cloaked. 

Starting from \cref{equ SM: Hamiltonian with Purcell filter,equ SM: Lindblad equation with Purcell filter} for the case of the transmon, i.e. $\hat H_\text{q} = \hat H_\text{tr}$ and $\hat H_g = ig \hat n_\text{tr} (\hat a^\dag - \hat a)$, we first diagonalize the Hamiltonians of the readout and Purcell cavities. In this basis, they become hybridized. Since $\omega_r \sim \omega_f \gg |J|$, we use the rotating-wave approximation in the cavity-cavity coupling. Moreover, adding intrinsic transmon decay and dephasing, we arrive at
\begin{equation}
\begin{split}
\partial_t \hat \rho =& -i[\hat H(t) + \hat H_\text{gate}(t), \hat \rho] \\
&+ \kappa_f \mathcal D[\sin\frac{\theta}{2} \hat a_+ + \cos\frac{\theta}{2}\hat a_-]\hat \rho \\
& + \gamma \mathcal D[\hat d]\hat \rho  + 2 \gamma_\phi \mathcal D[\hat d^\dag \hat d]\hat \rho,
\end{split}
\end{equation}
with $\hat H_\text{gate}$ given in \cref{equ SM: gate Hamiltonian}, $\hat d = \sum_j \sqrt{j} \ket{j}\bra{j+1}$ is a lowering operator in the energy eigenbasis of the transmon, $\hat H_\text{tr} \ket{j} = E_j \ket{j}$, and
\begin{equation}
\begin{split}
\hat H(t) =& \hat H_\text{tr} + \omega_+ \hat a^\dag_+ \hat a_+ + \omega_- \hat a_-^\dag \hat a_- \\
&+ig\hat n_\text{tr}\cos(\theta/2)(\hat a_+^\dag - \hat a_+) \\
&- ig\hat n_\text{tr}\sin(\theta/2)(\hat a_-^\dag - \hat a_-) \\
&+ i\mathcal{E}_1(t)\sin(\theta/2)(\hat a_+^\dag - \hat a_+) \\
&+ i\mathcal{E}_1(t)\cos(\theta/2)(\hat a_-^\dag - \hat a_-),
\end{split}
\end{equation}
where
\begin{equation}
\begin{pmatrix}
\hat a_+ \\ \hat a_-
\end{pmatrix} = \begin{pmatrix}
\cos(\theta/2) & \sin(\theta/2) \\
-\sin(\theta/2) & \cos(\theta/2)
\end{pmatrix} \begin{pmatrix}
\hat a \\ \hat f
\end{pmatrix},
\end{equation}
with $\tan \theta = 2J/(\omega_r - \omega_f)$ and 
\begin{equation}
\omega_{\pm} = \frac{\omega_r + \omega_f}{2} \pm \frac{1}{2}\sqrt{(\omega_r - \omega_f)^2 + 4J^2}.
\end{equation}

\begin{table}
\begin{tabular}{|c | c|} 
 \hline
 Parameter & Value \\ [0.5ex] 
 \hline\hline
$E_C/2\pi\hbar$ & 208.09 MHz \\
$E_J/2\pi\hbar$ & 16.23 GHz \\
$\omega_r/2\pi$ & 7.64744 GHz \\
$\omega_f/2\pi$ & 7.63166 GHz \\
$g/2\pi$ & 166.85 MHz \\
$J/2\pi$ & 26.14 MHz \\
$\kappa_f/2\pi$ & 29.1 MHz \\
$\gamma/2\pi$ & 6.35 kHz \\
$\gamma_\phi/2\pi$ & 18.04 kHz \\ 
$t_g$ & 25 ns \\
$\varepsilon_g/2\pi$ & 29.192 MHz \\
$d_g$ & 0.0869 \\
$\phi_g/2\pi$ & -0.0273 \\ [1ex] 
 \hline
\end{tabular}
\caption{Parameters for the numerical simulation of the X gate average error including a Purcell cavity.}
\label{tab: Parameters gate simulation}
\end{table}

The full set of simulation parameters that we use are shown in \cref{tab: Parameters gate simulation}. With these values of $\gamma$ and $\gamma_\phi$, we obtain a qubit lifetime and Ramsey coherence time of $T_1 = 25$ $\mu$s and $T_2=7.5$ $\mu$s, respectively, as in our experiment.

\begin{figure}[t!]
\includegraphics{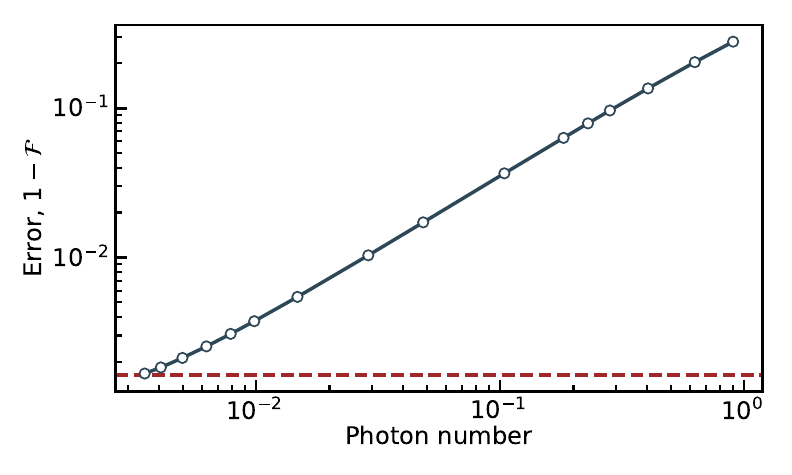}
\caption{\captiontitle{Predicted $X$-gate average error accounting for the Purcell cavity.} As a function of the number of photons in the readout cavity at the end of the gate, the blue curve with dots shows the error without cancellation, and the dashed line the error with cancellation. The minimum error attained is $\sim 0.163\%$. Parameters are indicated in \cref{tab: Parameters gate simulation}.}
\label{fig SM: theory gate with Purcell}
\end{figure}

For the numerical simulations of the X gate, we work in the displaced frame where the drive on the Purcell cavity is effectively passed to the qubit. This allows us to integrate the master equation with truncated Hilbert spaces for the modes $\hat a_\pm$ that are smaller than would be required if we worked in the laboratory frame. In the results shown in \cref{fig SM: theory gate with Purcell} (and Fig.~5 in the main text), we use the definition of \cref{equ SM: average gate fidelity}, where the six cardinal states of the Bloch sphere are now those of the logical subspace $\{\ket{\overline{g,0_+, 0_-}}, \ket{\overline{e, 0_+, 0_-}}\}$, corresponding to the ground and excited states of the qubit dressed by the hybridized modes $\hat a_\pm$ (here $0_\pm$ is the zero Fock state of these modes).

The gate error obtained from numerical simulations quantitatively agree with experiments in the presence of cloaking (see dashed red line in Fig.~5 in the main text and \cref{fig SM: theory gate with Purcell}). In the absence of cloaking, full blue line in \cref{fig SM: theory gate with Purcell}, we find numerically the expected increase of the gate error with cavity photon number but the agreement with the experimental observations is not quantitative.  While the average gate error in the presence of cloaking is coherence limited and thus simple to fit to the experimental results by adjusting $\gamma$ and $\gamma_\phi$, in the absence of cloaking the error depends on the quality of the fit of the bare parameters used in the model, as well as on a precise calibration of the experimental attenuation of drive power.  An accurate comparison of numerical results to experiment data as a function of photon number is therefore challenging. For this reason, we do not show the numerical results together with the experimental data  in Fig.~5 of the main text.

\section*{Supplementary Note 2 -- Experiment}

\subsection{Calibration of the cancellation tone}

\begin{figure}[t!]
    \centering
    \includegraphics{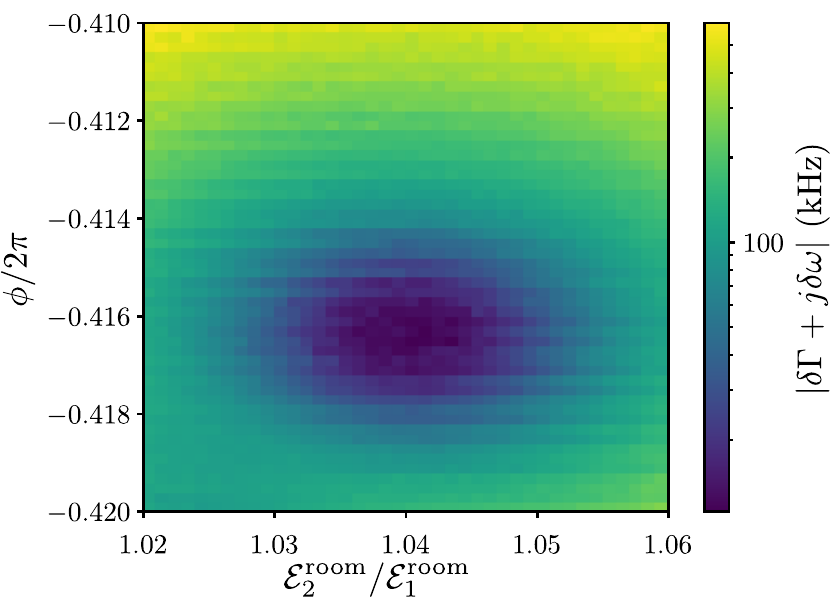}
    \caption{\captiontitle{Calibration of the cancellation phase $\phi$ and amplitude $\mathcal{E}_2^\text{room}$.} We minimize the extra-dephasing $\delta\Gamma$ and ac-Stark shift $\delta\omega$ measured by Ramsey interferometry. The cavity drive amplitude is $\varepsilon_1/2\pi = \SI{107}{MHz}$.}
    \label{fig:SI_calib_cancellation}
\end{figure}

The cancellation drive amplitude and phase is calibrated using Ramsey interferometry where the drives $\mathcal{E}_{1/2}$ are applied between the two $\pi/2$ pulses of the Ramsey sequence. For a given drive $\mathcal{E}_1$ on port 1, we minimize the extra-dephasing and ac-Stark shift $|\delta\Gamma+i\delta\omega |$, see \cref{fig:SI_calib_cancellation}. We used a cancellation tone ansatz of the form given in \cref{equ SM: approximate cancellation tone for a constant cavity drive}.

The Ramsey sequence (with a constant $\varepsilon_1$) is not the most sensitive way to measure a miscalibration of the cancellation tone in the transient regime. For example, with $\kappa/2\pi = \SI{10}{MHz}$ in the experiment reported here, the steady state of the cavity field is reached in $\sim \SI{100}{ns}$, which is small in comparison to the tens of microseconds of the employed Ramsey sequence. If further precision at early times is necessary, only two more calibration steps are needed to calibrate the time-delay mismatch and phase between the drive term at $\omega_1$ and the term at $\omega_r$ in \cref{equ SM: approximate cancellation tone for a constant cavity drive}. These calibrations can be done separately from the previous calibrations and do not influence the already calibrated parameters.

Nevertheless, as shown in \cref{fig SM: theory gates}e, corresponding to the worst-case scenario where the cancellation-tone term at $\omega_r$ is missed altogether (c.f. \cref{equ SM: anzat cancellation tone}), only a small error in the X gate is introduced.

\subsection{Histograms and overlap error}
\begin{figure}[h!]
    \centering
    \includegraphics{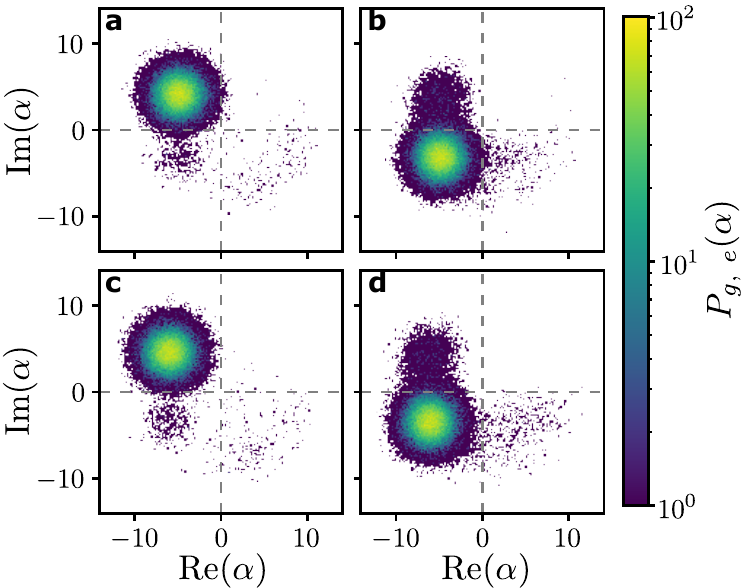}
    \caption{\captiontitle{Measured Husimi distributions for standard dispersive and arm-and-release readout.} Histograms of the demodulated heterodyne signal for $t_\mathrm{int}=\SI{196}{ns}$ and $\varepsilon_1/2\pi = \SI{63.7}{MHz}$ for the standard dispersive readout (\textbf{a} and \textbf{b}) and for the arm-and-release readout (\textbf{c} and \textbf{d}) when the qubit is prepared in its ground state (\textbf{a} and \textbf{c}) or in its excited state (\textbf{b} and \textbf{d}).}
    \label{fig:SI_IQ_blobs}
\end{figure}

Each heterodyne signal is demodulated over an integration time $t_\mathrm{int}$ resulting in a complex value $\alpha$. Repeating this measurement, we obtain probability distributions of the complex amplitudes $P_{g,e}(\alpha, t_\mathrm{int})$ when the qubit has been prepared in state $|g\rangle$ or $|e\rangle$. Typical probability distributions  are shown in \cref{fig:SI_IQ_blobs} for $t_\mathrm{int}= \SI{196}{ns}$ in case of standard dispersive or  arm-and-release readout. The measured histograms reveal two Gaussian distributions corresponding to each qubit state. Some readout amplitudes $\alpha$ fall out of these two distributions, which happens with a probability $p_\mathrm{out}= 0.07\%$, likely due to transmon ionization~\cite{Shillito2022}. The readout amplitude has been chosen on the onset of ionization in order to reach a good trade off between the error of finite separation between pointer states and the error $p_\mathrm{out}$ of ionizing the transmon.

From these histograms $P_{g,e}(\alpha, t_\mathrm{int})$, we compute the overlap (their normalized 2D scalar product)
\begin{equation}
O(t_\mathrm{int}) = \frac{\int P_g(\alpha, t_\mathrm{int}) P_e(\alpha, t_\mathrm{int}) d\alpha}{\sqrt{\int P_g(\alpha, t_\mathrm{int})^2 d\alpha}\sqrt{\int P_e(\alpha, t_\mathrm{int})^2 d\alpha} }.
\end{equation}
To compute the average fidelity $\mathcal{F} = 1 - [P(g|e)+P(e|g)]/2 $ with $P(x|y)$ the probability to measure $x$ when state $y$ was prepared, we project the dataset onto the imaginary axis. Defining the threshold $q_\mathrm{th} = 0$, we compute the error probabilities $P(g|e)= P(\Im(\alpha) > q_\mathrm{th}|e) =\SI{1.07\pm 0.14}{\%}$ and $P(e|g)= P(\Im(\alpha) \leq q_\mathrm{th}|g)=\SI{0.23\pm 0.14}{\%}$. We thus obtain $\mathcal{F} = \SI{99.35\pm 0.14}{\%}$. The uncertainty $\pm\SI{0.14}{\%}$ comes from the finite number $N=10^6$ of repetitions. 
The error of $\SI{0.65}{\%}$ in the average infidelity is mostly explained by the following processes. By fitting the distribution with double Gaussians~\cite{Walter2017}, we compute the error due to finite separation ($\sim 0.13\%$) between the two Gaussians corresponding to the two states of the qubit.
In these experiments, the qubit is first prepared in the ground state $\ket{g}$ using measurement-based feedback with the usual dispersive readout. Using a second standard dispersive readout, we estimate the error of wrong preparation in the ground state before the arming step as $P_\mathrm{disp} (e|g) \sim 0.2\%$. Incorrect excited state preparation is also explained by imperfect $\pi$ pulse due to the coherence limit of the qubit (giving in average, an error of $\sim 0.08\%$, see Fig.~5 of the main text). Error due to relaxation during the measurement correspond to $\sim \frac{1}{2} (1 - e^{-t_\mathrm{int}/2T_1}) \sim 0.2\%$.

\subsection{Pre-arming amplitude and phase optimization}

Because the cavity does not respond at the same frequency during the arming and  release steps, the amplitude and phase of the arming step needs to be optimized to obtain the fastest separation between cavity states during readout. Indeed, during the arming step, the steady-state amplitude is given by $\alpha^\text{s}_\text{a} = (\varepsilon_{1,a} e^{i\phi_{1,a}}/2)/(\omega_r-\omega_1-i\kappa/2)$ while in the release step, it is given by $\alpha^\text{s}_\text{r,i} = (\varepsilon_{1,r} e^{i\phi_{1,r}}/2)/(\tilde \omega_r + \chi_i-\omega_1-i\kappa/2)$ and depends on the state $i$ of the qubit. 
For our parameters and choosing $\omega_1 = \tilde \omega_r +(\chi_g+\chi_e)/2$, driving at the same amplitude and phase leads to an amplitude ratio of $|\alpha^\text{s}_\text{a}|/|\alpha^\text{s}_\text{r}| = 0.58$ and phase difference $\phi_{\alpha^\text{s}_\text{a}}-\phi_{\alpha^\text{s}_\text{r}} = 0.154 \times 2\pi$ where $\phi_{\alpha^\text{s}_\text{r}}= (\phi_{\alpha^\text{s}_\text{r,g}}+\phi_{\alpha^\text{s}_\text{r,e}})/2$ is the average phase obtained for the two states of the qubit.

\begin{figure}[t!]
    \centering
    \includegraphics{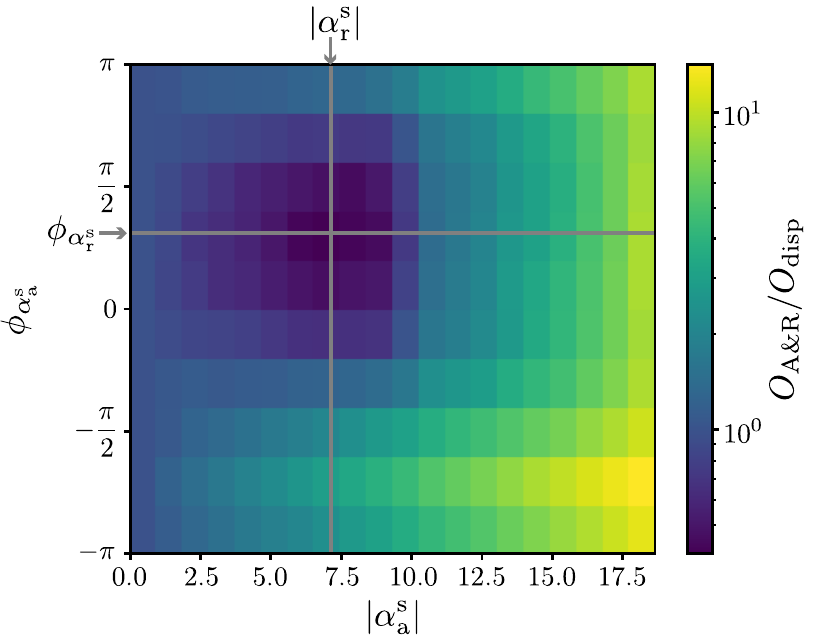}
    \caption{\captiontitle{Readout overlap error comparison for standard disperive and arm-and-release methods.} Error ratio $O_\text{A\&R}/O_\text{disp}$ between arm-and-release $O_\text{A\&R}$ and standard dispersive $O_\text{disp}$ readout errors as function of arming amplitude $|\alpha_\text{a}^\text{s}|$ and phase $\phi_{\alpha_\text{a}^\text{s}}$. The overlap errors are extracted for a 140 ns integration time. }
    \label{fig:SI_A&R_optimisation}
\end{figure}

To optimize the readout for the arm-and-release protocol, we measure the overlap error ratio $O_\mathrm{A\&R}/O_\mathrm{disp}$ between the arm-and-release $O_\mathrm{A\&R}$ and the standard dispersive $O_\mathrm{disp}$ readouts (\cref{fig:SI_A&R_optimisation}). The chosen arming amplitude and phase is the one minimizing the error ratio $O_\mathrm{A\&R}/O_\mathrm{disp}$.

\subsection{Randomized benchmarking under cavity drives}

\begin{figure}[t!]
    \centering
    \includegraphics{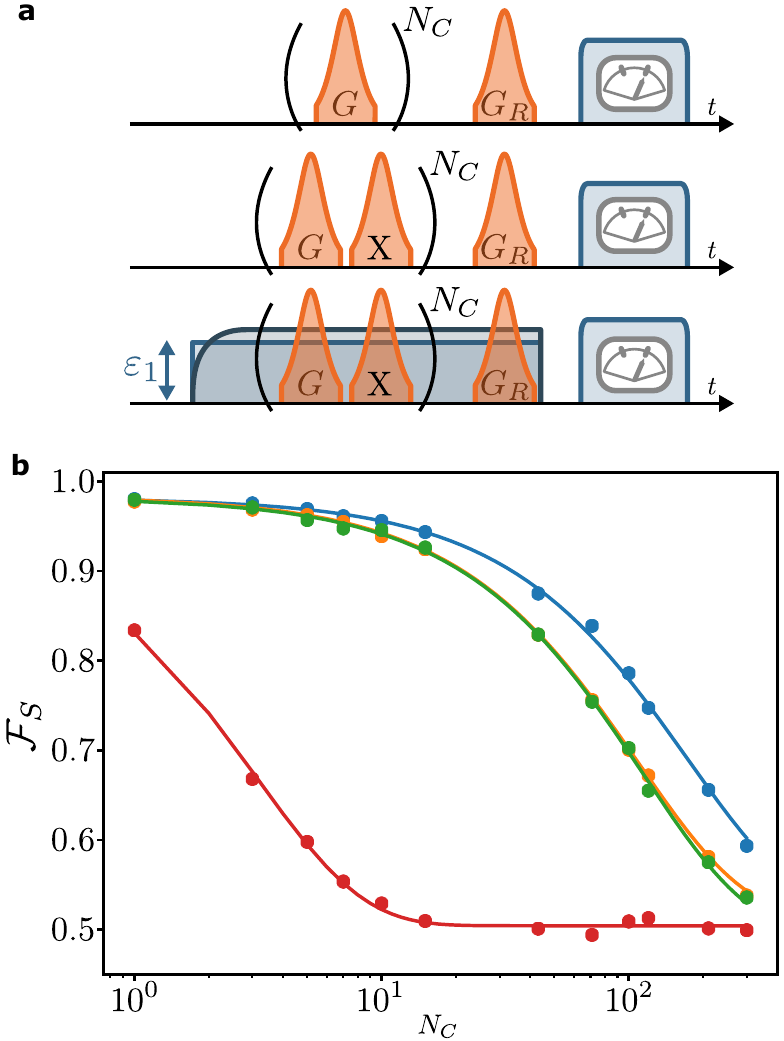}
    \caption{\captiontitle{X gate randomized benchmarking.} \textbf{a}, Pulse sequences used for the randomized benchmarking leading to Fig.~5 of the main text: Top, reference sequence; middle, interleaved sequence; bottom, interleaved sequence under arming drives. \textbf{b}, Dots: measured sequence fidelity (probability to end in the same state as the initial one) as a function of circuit depth $N_C$. Lines: fits using $\mathcal{F}_S = A p_{G}^{N_C} + B$. Blue: reference sequence, orange: interleaved without drives, red: interleaved with $\mathcal{E}_1$ drive only, and green: interleaved with $\mathcal{E}_1$ and $\mathcal{E}_2$ drives. The drive $\varepsilon_1/2\pi = \SI{6.4}{MHz}$ corresponds to 0.37 photons on average without cancellation and 0.14 photons with cancellation.}
    \label{fig:SI_RB}
\end{figure}

The gate errors under drives in Fig.~5 of the main text are estimated using randomized benchmarking~\cite{Knill2008}. For that purpose, we compare the fidelity of different pulse sequences (\cref{fig:SI_RB}a). In a reference pulse sequence, a number $N_C$ of random Clifford gates is applied, followed by a recovery gate before reading out the qubit state. The sequence fidelity $\mathcal{F}_S$ is fitted using $\mathcal{F}_S = A p_{G, \mathrm{ref}}^{N_C} + B$ from which we extract $p_{G, \mathrm{ref}}$. The same fitting procedure is applied for an interleaved sequence where the gate under test is interleaved with the random gates, resulting in the probability $p_{G}$ (see \cref{fig:SI_RB}b).
The average gate error is then extracted as $\epsilon_X = \frac{1}{2} (1 - \frac{p_{G}}{p_{G, \mathrm{ref}}})$.

\subsection{Sample and measurement setup}

The large features of the sample are made by optical lithography on a Tantalum thin film on a Sapphire substrate, while the Josephson junction of the transmon qubit is fabricated via electronic lithography followed by angle deposition of Al/AlOx/Al in a Plassys evaporator. The readout mode is a $\lambda/4$ coplanar waveguide resonator. The Purcell filter is also a $\lambda/4$ coplanar waveguide resonator, inductively coupled to the readout mode, and is used as a bandpass filter around the readout frequency.

\begin{figure*}
    \centering
    \includegraphics[width=14cm]{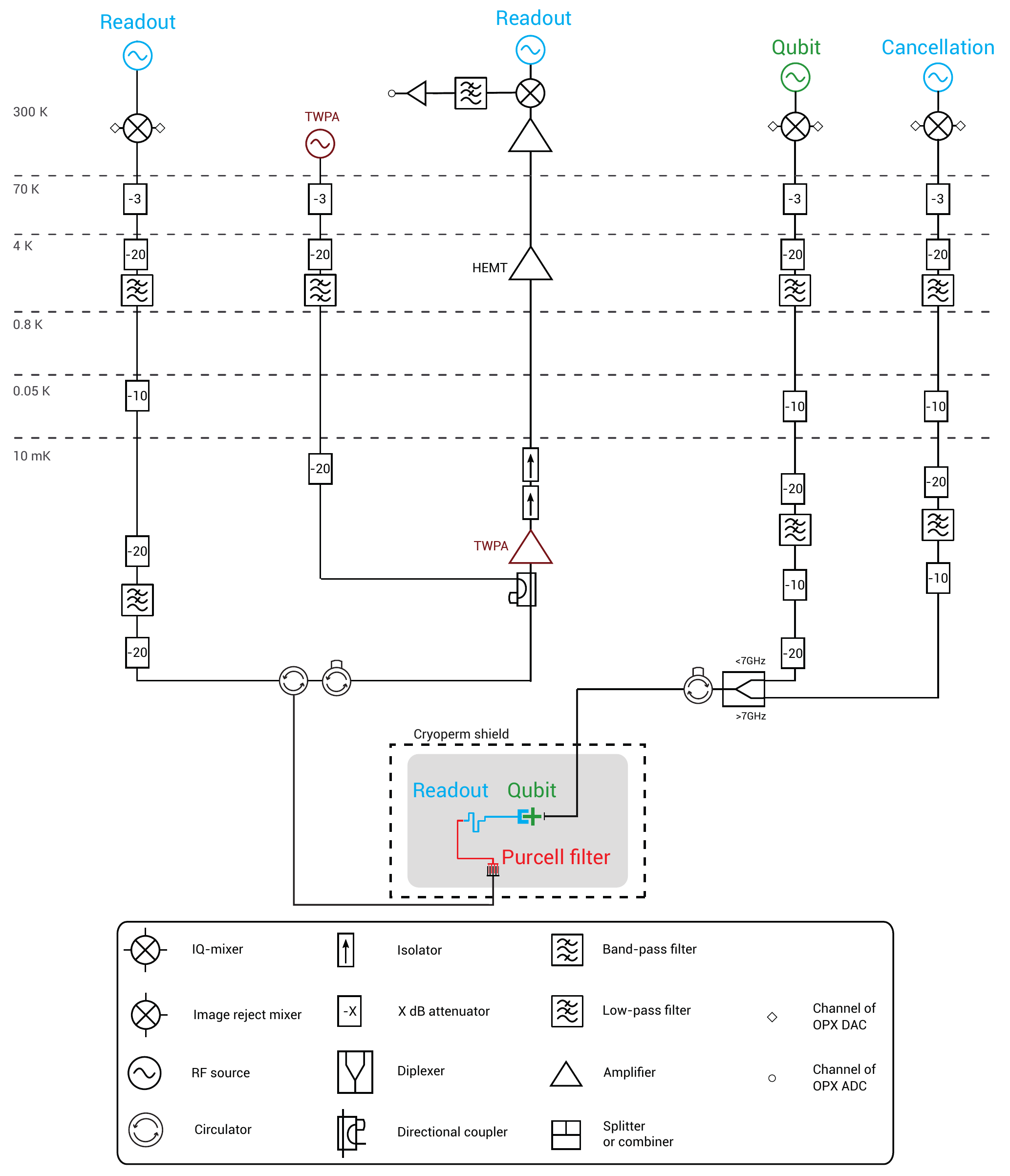}
    \caption{\captiontitle{Schematic of the microwave setup.}}
    \label{fig:microwave_setup}
\end{figure*}

The sample is cooled down to \SI{10}{mK} in a dilution refrigerator. The diagram of the microwave wiring is given in \cref{fig:microwave_setup}. The qubit, readout and cancellation pulses are generated by modulation of continuous microwave tones produced respectively by generators Anapico APSIN20G and Anapico APSIN12G (readout and cancellation tones use the same local oscillator). They are modulated via IQ-mixers where the intermediate frequency (a few tens of \si{MHz}) modulation pulses are generated by 6 channels of an OPX from Quantum Machines with a sample rate of \SI{1}{GS\per \second}. The acquisition is performed, after down-conversion by its local oscillator, by digitizing the \SI{100}{MHz} signal with the \SI{1}{GS\per \second} ADC of the OPX. The qubit and cancellation pulses are multiplexed into a single transmission line using a diplexer at the lowest temperature stage.

\section*{Supplementary Note 3 -- Potential limitations}
\label{sec: Potential limitations}

\subsection{The need of two driving ports}

As previously indicated, qubit cloaking necessitates distinct driving ports---one to address the nonlinear mode and another to address the cavity. In the case of circuit QED devices, this requirement is typically fulfilled with the presence of a port for logical operation on the qubit and a second port for readout on the measurement cavity. In other cavity-based platforms, a single driving port may be easily accessible. With distinct driving ports, a possible limitation is that the optimization---in terms of attenuation of thermal noise, dynamical range, or filtering---of the input transmission line for its use in driving the qubit might compete with its optimization for its use for cloaking. This is particularly true because of the requirement to drive the qubit port at the cavity frequency, and relatively large drive amplitudes necessary for cloaking.

\subsection{Qubit cloaking in the ultrastrong coupling regime}

In the ultrastrong coupling regime, where the qubit-cavity coupling is comparable to the cavity and qubit transition frequencies, the dissipator no longer assumes the form in \cref{eq:SimpleME} but rather takes a correlated qubit-cavity form~\cite{Rivas2010, Beaudoin2011}. The assumption of separate dissipation channels for the qubit and cavity is no longer a good approximation and the cloaking drive on the qubit cannot cancel the incoherent effects induced by this correlated dissipation.

\subsection{The qubit cannot be protected from the thermal excitations in the cavity}

While the qubit can be cloaked from the coherent drive on the cavity even in the presence of thermal excitations in the cavity, it cannot be cloaked from the cavity thermal excitations themselves: a coherent tone cannot cloak an incoherent excitation.

\bibliography{references}